\documentclass{JHEP3}
\usepackage{epsfig}
\newcommand{\be}{\begin{equation}}
\newcommand{\ee}{\end{equation}}
\newcommand{\bea}{\begin{eqnarray}}
\newcommand{\eea}{\end{eqnarray}}
\newcommand{\IR}{\mathbb{R}} 

\def\IZ{\relax\ifmmode\hbox{Z\kern-.4em Z}\else{Z\kern-.4em Z}\fi}
\newcommand{\IS}{{\bf S}}

\newcommand{\non}{\nonumber \\}

\def\half{{1 \over 2}} 
\def\tr{{\rm tr}}
\def\del{{\partial}}
\def\room{~\rule[-2mm]{0mm}{8mm}}

\def\tlt{\tilde{t}}
\def\tr{\tilde{r}}
\def\tz{\tilde{z}}

  \def\eps{\epsilon}

\def\trho{{\tilde \rho}}

\def\room{~\rule[-2mm]{0mm}{8mm}}
\def\presub{\vspace{.5cm} \noindent}

\def\bi{\begin{itemize}} \def\ei{\end{itemize}}

\def\({\left(} \def\){\right)}
\def\[{\left[} \def\]{\right]}
\preprint{{\tt hep-th/0502033}}

\title{ \center{Choptuik Scaling and The Merger Transition}}

\author{
Barak Kol \\
 Racah Institute of Physics\\
 Hebrew University \\
 Jerusalem 91904,
 Israel\\
{\tt barak\_kol@phys.huji.ac.il}}

\abstract{The critical solution in Choptuik scaling is shown to be
closely related to the critical solution in the black-string
black-hole transition (the merger), through double analytic
continuation, and a change of a boundary condition. The interest in
studying various space-time dimensions $D$ for both systems is
stressed. Gundlach-Hod-Piran off-critical oscillations, familiar in
the Choptuik set-up, are predicted for the merger system and are
predicted to disappear above a critical dimension $D^*=10$. The
scaling constants, $\Delta(D),\, \gamma(D)$, are shown to combine
naturally to a single complex number.}

\begin{document}

\begin{flushright}
\emph{To my brothers,  \hspace*{2.3cm} \\
 Boaz, Amos, Amir and Yohai}.
\end{flushright}

\section{Introduction}
\label{introduction}

This work describes a correspondence between two gravitational
systems: Choptuik scaling and the ``merger'' transition in the
black-hole black-string system.

Choptuik scaling \cite{Choptuik} describes the famous phenomenon
observed (in numerical simulations) at the threshold for black
hole production in a spherically symmetric gravitational collapse.
At threshold, also known as criticality, the solution approaches
an attractor solution as one approaches the space-time point where
the black hole is ``marginally'' formed. The observed independence
of initial conditions (as long as one tunes one parameter for
criticality) is known as ``universality''. The solution has
discrete self-similarity, known as ``echoing'', with scaling
constant denoted by $e^\Delta$, and it exhibits a ``critical
exponent'', $\gamma$. See also the review \cite{GundlachRev} and
references therein.

The merger transition originates in the black-string black-hole
transition, which occurs whenever extra compact dimensions are
present (see \cite{TopChange}, the review \cite{review} and
references therein). Instead of analyzing the full time-evolution
during phase transition it suffices, for purposes of determining
the end-state, to consider only stable static solutions, and it
turns out to be convenient to consider unstable static solutions
as well. For ``phase conservation'' reasons it was first predicted
\cite{TopChange} and recently numerically confirmed
\cite{KudohWiseman2} that there exists a path of solutions joining
the branch of increasingly non-uniform black-strings with the
black hole branch. Locally, at the point of minimal horizon
radius, or ``waist'', a topology changing transition occurs, where
not only the horizon topology changes, but actually the manifold
topology changes as well (at least in the Euclidean, Wick rotated
solutions, gotten from the static solutions). This transition was
called ``merger'', since it can be thought to describe the merger
of a large enough black-hole into a black-string.

After reviewing the Choptuik and merger systems in section
\ref{review}, it is shown in section \ref{correspondence} that the
two are closely related, as anticipated in \cite{TopChange} (p.21
bottom of page), and more recently in \cite{GIR}\footnote{I thank
Nissan Itzhaki for introducing me to the observation made in that
paper.} in connection with double analytic continuation (see
\cite{MSY,JMS} for related discussions) of D-branes.
 In both systems space-time is effectively 2d after accounting for symmetry,
and once the Choptuik scalar in $d$ dimensions is interpreted as
arising from a Kaluza-Klein reduction in $D=d+1$ dimensions, they
are seen to have the same matter content. Moreover, I show that
once one performs \emph{a double analytic continuation} the two
systems have precisely the same action. It should be noted
however, that the two analytic continuations are of a different
character: one is trivial in the sense that the fields do not
depend on the rotated coordinate, while the other is non-trivial,
involving an essential coordinate, one which the fields depend on,
and the success of the rotation (reality of the solution) relies
on the fields being even in that coordinate.

In order for the solutions to correspond under double analytic
continuation, it is not enough that the actions coincide (and
therefore the equations of motion) but the boundary conditions
(b.c.) must correspond as well. In both cases we are seeking a local
solution near a point-like singularity. In Choptuik it is the
marginal black hole and in the merger it is the marginally pinched
horizon. Locality means that all scales are forgotten near the
singularity and thus scale periodicity is a common b.c.
Alternatively, the periodic b.c. may be replaced by the attractor
mechanism where in both cases it is necessary to fine-tune one
parameter -- in Choptuik it is the initial imploding wave while in
the merger it is a b.c. such as the temperature that parametrizes
the curve of solutions. These are boundary conditions along the
``scaling direction'', but we still need b.c. along the
``tangential'' direction. There one actually finds two kinds of
b.c., and in this respect the two systems differ.

\begin{figure}[t!]
\centering \noindent
\includegraphics[width=7cm]{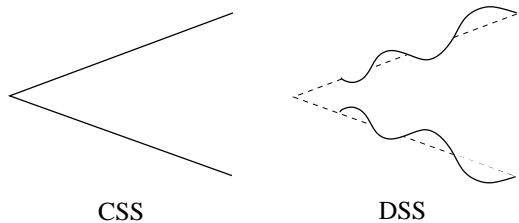}
\caption[]{An illustration of a continuously self-similar geometry
(CSS), which we often call a cone, and a discretely self-similar
geometry (DSS).} \label{cone-illus}
\end{figure}

Self-similar solutions, such as the critical solutions discussed
here, can be either Continuously Self-Similar (CSS) or Discretely
Self-Similar (DSS). CSS solutions return to themselves after
rescaling by any constant and a cone is a good mental picture for
them, while DSS solutions are invariant only by a rescaling by a
specific rescaling factor (and its power) and can be represented by
a wiggly cone with logarithmically periodic wiggles (see figure
\ref{cone-illus}). In subsection \ref{cones-GHP} we discuss CSS
solutions and especially the evidence \cite{AKS} for the double-cone
(which is CSS) being the critical merger solution. That leads to an
interpretation of the complex exponents that appear in the
perturbations of the double cone \cite{TopChange} as critical
exponents of the merger system. The real part is predicted to be
related to the critical exponent $\gamma$, which sets the dimensions
of the off-criticality parameter $(p-p^*)$, while the imaginary part
is predicted to be related to the critical exponent $\Delta$. In
critical collapse, on the other hand, the critical solution is DSS.
While there $\Delta$ manifests itself as the log-periodicity of both
the critical solutions and the Gundlach-Hod-Piran (GHP) oscillations
of off-critical quantities, for the merger we predict the latter
without the former.  Altogether, the scaling constants $\gamma,\,
\Delta$ are shown to combine naturally into a single complex number
related to the perturbative exponents, the precise relation being
(\ref{complexify}), which generalizes the well-known connection of
$\gamma$ with the linearized analysis (\ref{gamma-lambda}).

The differences between the merger system and the Choptuik critical
collapse are summarized in table \ref{table1}. Some implications are
discussed in section \ref{consequences}. Briefly, they are \bi
 \item A prediction of GHP oscillations in the merger system.

 \item The cone provides a prediction of the critical exponents
$\Delta,\,\gamma$  and a critical dimension $D^*=10$ for the merger
(\ref{gamma-pred2}-\ref{gamma-pred3}).

 This prediction has some analogy with \cite{Frolov4,Frolov2} which
analytically estimate the critical exponents of Choptuik scaling by
analyzing perturbations around the CSS Roberts solution
\cite{Roberts} in 4 and higher dimensions, respectively.

 \item Perhaps there are similarities between the solutions and scaling constants of the two systems
as they differ only by a change of b.c. and therefore perhaps some
results would carry over from the merger to the standard Choptuik
system.
 \ei

\begin{table}[t!]
\centering
\begin{tabular}{l||c|c}
    & \hspace*{.5cm} Merger \hspace*{.5cm} &  Choptuik Critical Collapse \\
 \hline \hline
  (Essential) Signature  &  Euclidean & Lorentzian \\
  Boundary condition &  Reflection & Axis regularity  \\
  Self Similarity of critical solution &  Continuous\footnote & Discrete  \\
  %\footnote{Direct evidence is lacking so far.}
\end{tabular}
\caption[]{A summary of the differences between the critical merger
solution and the critical Choptuik solution.} \label{table1}
\end{table}

\footnotetext[\value{footnote}]{Direct evidence is lacking so far.}

\noindent {\bf Distant outlook}. Finally, I would like to discuss
some general but non-rigorous lessons \bi

 \item Choptuik scaling is well-known to be very similar to
conformal field theories, as it exhibits scale invariance and
critical exponents. Equipped with the modern perspectives of
holography, and the duality between 2d gravity and matrix models,
I find it suggestive to predict that quantum gravity near the
singularity (for both the spherical collapse and the merger) is
described by some (yet unknown) large $N$ \emph{conformal matrix
model}.

\item The time evolution during the Gregory-Laflamme decay
 inevitably leads to a pinching singularity with high
energy effects, irrespective of the (low energy) initial
conditions. It is very probable that this singularity would
exhibit scaling as well. In that sense it is a manifestation of
\emph{self-organized criticality}. It would be interesting to
understand the critical phenomena near this singularity which
distinguishes itself by involving three essential coordinates --
$(r,z,t)$, rather than two (see \cite{CLOPPV} for a simulated time
evolution).

 \ei

\presub {\bf Note added} (version 3). The perspective gained by
developments including \cite{SorkinOren,ItzhakiPrivat,AKS} proves
the main results to be \bi
 \item The relation between the merger and Choptuik scaling through double analytic
continuation (section \ref{correspondence}).
 \item  The prediction of GHP oscillations
for the merger system in subsection \ref{cones-GHP} with critical
exponents given by (\ref{gamma-pred2}-\ref{gamma-pred3}). \ei

Moreover, following the same developments some of the questions
raised in the original ``implications'' section were answered as we
proceed to describe. The body of the paper, sections
\ref{review},\ref{correspondence} are mostly unchanged, while
sections \ref{introduction},\ref{consequences} were changed
accordingly. \bi

\item The suggestion in version 1 (v1) to study Choptuik's
critical collapse in various dimensions was taken up successfully
in \cite{SorkinOren} (see section \ref{consequences} for a
discussion).

\item The idea in v1 to numerically study the analytical
continuation of either merger or spherical collapse was conditioned
on the stability of the analytically continued solutions. I came to
believe that these continuations are unstable and so would be any
numerical implementation of them \cite{ItzhakiPrivat}. Instabilities
are typical in the analytic continuation of actions -- the essential
argument can be seen in the harmonic oscillator Lagrangian $L \sim
\dot{x}^2-x^2$. If one performs an analytic continuation $t \to i\,
t$ the kinetic term is inverted, and after a multiplication of $L$
by an overall minus sign we find that the potential term is inverted
leading to instability.

\item \cite{AKS} showed that the double cone is an attractor at
codimension 1 for a certain class of admissible perturbations and
that was interpreted as significant evidence that it is the critical
merger solution. Therefore the critical solution is probably CSS,
rather than DSS as originally predicted in v1. The cause for the
error was that originally, the off-critical oscillations of the
merger were taken to imply DSS, just like in the critical collapse
the two measure the same log-period $\Delta$. However, it turns out
that the implication is only in the converse direction, namely DSS
implies GHP oscillations, and the latter can exist even without the
former. In particular, the perturbations are not interpreted here as
tachyons anymore. Subsection \ref{cones-GHP} was changed
accordingly.

\ei

\newpage
\section{Review}
\label{review}

We start by reviewing the two concepts to be linked in this work.

\subsection{The merger transition}

As reviewed in the introduction the merger transition originates
in the black-string black-hole transition, which occurs whenever
extra compact dimensions are present (see \cite{TopChange}, the
review \cite{review} and references therein). The merger
transition is the local topology change at the ``waist'' as one
moves from an unstable black-string to an unstable black-hole.

\begin{figure}[t!]
\centering \noindent
\includegraphics[width=7cm]{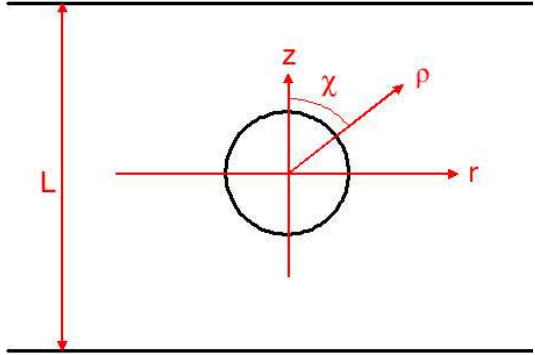}
\caption[]{Definition of coordinates for the merger system. For
backgrounds with a single compact dimension the essential geometry
is 2d and Euclidean after suppressing the time $t$ and angular
coordinates in the extended dimensions.  The cylindrical
coordinates $(r,z)$ are defined such that $z \sim z+L$ is the
coordinate along the compact dimension and $r$ is the radial
coordinate in the extended spatial directions. Locally at the
``pinching singularity'' we define another set of local
coordinates $(\rho,\chi)$ (defined only for $\rho \le L/2$), which
are radial coordinates in the 2d plane with origin at the
singularity. We shall sometimes call $\rho$ a ``scaling
coordinate'' and $\chi$ ``tangential''.} \label{coordinates}
\end{figure}

\noindent {\bf The system}. One considers black objects in
 a flat $D$-dimensional space-time background $\IR^{D-2,1} \times
\IS^1$, where the compact coordinate is denoted by $z$ and its
length is $L: ~z \sim z+L$ (see figure \ref{coordinates}). The
matter content is pure gravity and the action is the standard
Einstein-Hilbert action $S_{D}=1/(16\pi G)\, \int R\, \sqrt{g}
d^Dx$.

The black objects are spherical and static, namely the isometry is
$SO(D-2)_\Omega \times U(1)_t$ (``static'' means also time
reversal symmetry). The most general metric consistent with this
symmetry is \be \label{general-ansatz}
 ds^2 = -e^{2\, A}\, dt^2 + ds^2_{(r,z)} + e^{2\, C}
 d\Omega^2_{D-3} ~,\ee
where all fields are defined on the Euclidean $(r,z)$ plane,
$ds^2_{(r,z)}$ is an arbitrary metric on the plane and since the
metric is static an analytic continuation $t \to it$ is trivial
and we may freely switch between the Euclidean and Lorentzian
signatures.

Altogether the problem is defined in the Euclidean $(r,z)$ plane
and the field content is a 2d metric and two scalars $A,\, C$.
That means that we can write down a 2d action for these fields
without loosing any of the equations of motion. The action is \bea
 S &=& {\beta\, L \over 4\, G_D}\, \int dV_2 ~e^{A+2C} \cdot \non
 &\cdot& \left[R_2 + (D-3)(D-4)\, e^{-2C} + (D-3)(D-4)\, (\del C)^2 +
 2(D-3)\,(\del A)(\del C) \right] \label{general-action} \eea
where $R_2$ is the 2d Ricci scalar and $dV_2:= \sqrt{g_2}\, dr\,
dz$ is the volume element. The total number of fields is 5. Two
fields may be eliminated by a choice of coordinates in the plane
which leaves us with three fields.

\begin{figure}[t!] \centering \noindent
\includegraphics[width=9cm]{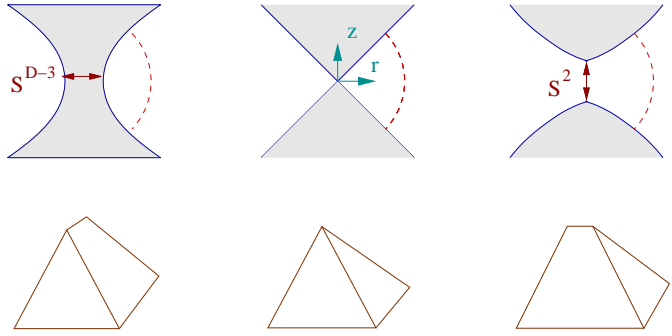}
\caption[]{The merger transition. A black string (left) turns into
a black hole (right) as the waist pinches. Shaded regions are
inside the horizon and the dashed line is a boundary far away. The
singular configuration is a cone over $\IS^2 \times \IS^{D-3}$ --
the double-cone.} \label{merger}
\end{figure}

\presub {\bf The double-cone}. Intuitively the transition from
black string to black hole involves a region where the horizon
becomes thinner and thinner as a parameter is changed until it
pinches and the horizon topology changes. This region is called
``the waist'' and this process is described in the upper row of
figure \ref{merger} using the $(r,z)$ coordinates defined in
figure \ref{coordinates}. It is important to remember that all
metrics under consideration are static and that they change as we
change an external parameter, not time.

A topological analysis \cite{TopChange} indicates that {\it the
local topology of (Euclidean) spacetime is changing}, not only the
horizon topology. Moreover, {\it the topology change can be modeled
by the ``pyramid'' familiar from the conifold transition} (see the
lower row of figure \ref{merger}). By the nature of topology, in
order to change it there must be at least one singular solution
along the way (with at least one singular point). The simplest
possibility, which is also realized in the conifold is  to assume
that the {\it the singular topology is the cone over} $\IS^{D-3}
\times \IS^2$, which we term {\it the ``double-cone''}.

It is easy to write down a Ricci flat metric for the singular
solution,  which is moreover continuously self-similar
(CSS).\footnote{We shall freely interchange the terms ``cone'' and
``CSS''.} The metric is \be
 ds^2 = d\rho^2 + {\rho^2 \over D-2}
 \left[  d\Omega^2_{\IS^2} + (D-4)\, d\Omega^2_{\IS^{D-3}}
 \right] ~, \label{double-cone-metric} \ee
where the $\rho$ coordinate measures the distance from the tip of
the cone, which is the only singular point, and the constant
pre-factors are essential for Ricci-flatness.

It turns out that the double-cones may have oscillating
perturbations and that their existence surprisingly depends on a
critical dimension $D^*=10$ \cite{TopChange}. The relevant mode is a
function $\eps(\rho)$ which inflates slightly one of the spheres
while shrinking the other. The ansatz for the perturbation is \be
\label{stab_ansatz_gen}
 ds^2 = d\rho^2 + {\rho^2 \over D-2}
 \left( e^{2\, \eps/2}\, d\Omega^2_{\IS^2}
  +(D-4)\, e^{-2\, \eps/(D-3)}\, d\Omega^2_{\IS^{D-3}} \right)~. \ee
{\it A priori} one could start with two separate scale functions,
one for each sphere, but a constraint relates them as above.

Considering the zero mode for $\eps$, namely linearized
deformations around the double-cone, one finds from the equations
of motion for the ansatz (\ref{stab_ansatz_gen}) \bea
 \eps &=& \rho^{s_\pm} \non
 s_\pm &=& {D-2 \over 2}\,\left(-1 \pm i
\sqrt{{8 \over D-2}-1}\right) ~. \label{linearized-exponents} \eea

The imaginary part, $\Im(s)$, causes the oscillations. For $D \ge
D^*:=10$ we see that $s_\pm$ become purely real, namely $D^*=10$ is
\emph{a critical dimension}.

\subsection{Choptuik scaling}

Consider the threshold for black hole production. It is a
co-dimension 1 surface (or ``wall'') in the space of initial
conditions of any gravitational theory. ``Choptuik scaling'' is
the term for the critical phenomena physics at this threshold.
Here we shall review the basic system where the famous discoveries
of Choptuik were made through computerized simulations of
spherical collapse \cite{Choptuik} and describe its salient
features. More information, and a survey of other systems can be
found in the excellent review \cite{GundlachRev}.

\presub {\bf The system}. One considers an implosion of a
spherical shell. There is a small price to pay for the high degree
of symmetry - the shell cannot be made of gravitational waves
(which do not possess an S-wave due to their spin 2). A simple
choice for the matter content is a single scalar field, $\Phi$.
Thus the action is taken to be \be
 S_{Choptuik }={1 \over 16 \pi\, G_N} \int \sqrt{-g}\, d^d
 x ~\( R + \half (\del \Phi)^2 \)
 \label{choptuik-action} \ee

\begin{figure}[t!]
\centering \noindent
\includegraphics[width=7cm]{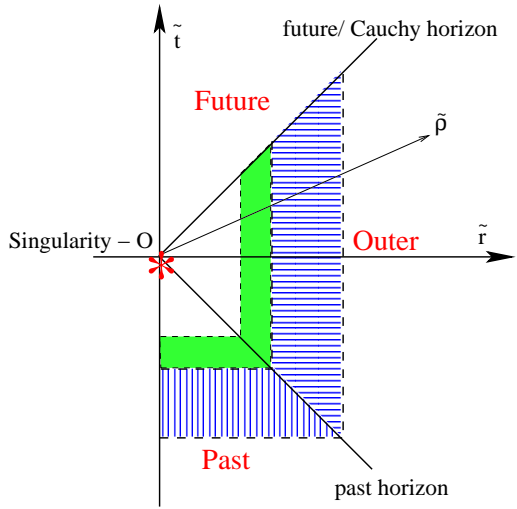}
\caption[]{Definition of coordinates for spherical collapse system
and Choptuik scaling (based on
\cite{Gundlach96,GarciaGundlachGlobal}). The essential coordinates
$(\tr,\tlt)$ parametrize a 2d Lorentzian plane after suppressing
the angular coordinates. The scaling direction is along lines of
fixed $\tlt/\tr$ and may be parametrized by
$\trho^2:=\tr^2-\tlt^2$. $\tz$ parametrizes an additional
dimension, the dimensional uplift of the scalar field $\Phi$. The
domain is made out of three patches: the past patch, bounded by
the $\tr=0$ axis and the past horizon, the outer patch bounded by
the past and future horizons and the future patch bounded by the
future (Cauchy) horizon and the axis. The critical solutions is
periodic on smaller and smaller scales as the singularity is
approached. One period is denoted by the line-filled (blue) region
and a second one is denoted by a shaded (green) regions. The
pattern continues towards the singularity.} \label{Choptuik-coord}
\end{figure}

Spherical symmetry means that the essential coordinates (upon
which the fields depend), $(\tr,\tlt)$ parametrize a 2d Lorentzian
plane (see figure \ref{Choptuik-coord}). We use tilded coordinates
for the Choptuik solutions to distinguish them from the untilded
coordinates for the merger.

Then one considers a family of initial conditions parametrized by
some parameter $p$. For instance, one could take a family of
Gaussian-profiled scalar waves, or any other profile, with $p$
being proportional to the initial amplitude. For small enough $p$
the linear approximation is valid, and by superposition the waves
go through the origin and ``reflect'' back to infinity. For large
enough $p$ a black hole forms. Thus, for any such family the
threshold of black hole formation defines a critical value of the
parameter which is denoted by $p_*$. Naturally, the value of $p_*$
depends on the chosen family.

\presub {\bf Main results}. It was found that this system displays
\emph{universality}, namely, some properties are independent of
the chosen family of initial conditions. There are two main
universal quantities \bi
 \item The critical exponent $\gamma$.
 \item The log-periodicity $\Delta$.
 \ei

\presub \emph{The critical exponent}. Consider the black-hole mass
as a function of $p$, namely $M_{BH}=M_{BH}(p)$. For $p<p_*$
$M_{BH}=0$ while for $p>p_*$ $M_{BH}>0$ and thus $p=p_*$ is a
non-analytic point of this function. It turns out that for $p
\gtrsim p_*$ \be
 M_{BH} \simeq (p-p_*)^\gamma \label{def-gamma} ~.\ee
 where $\gamma$ is a universal critical exponent. In 4d \cite{Gundlach96} \be
 \gamma \simeq 0.374 \label{gamma-4d} ~.\ee

\presub \emph{Echoing and log-periodicity}. For $p=p_*$ there is a
special point in space-time where the black-hole is ``just
almost'' being created. Clearly it is at $\tr=0$ and we might as
well shift $\tlt$ so that it has $\tlt=0$. Let us denote this
point by $O$. As the simulation approaches $O$ the solution starts
repeating itself on shorter scales and on shorter times. More
precisely the solution approaches the ``critical solution'',
$Z_*$, independently of initial conditions, and $Z_*$ is
discretely self-similar (DSS) with $O$ being its fixed-point.

In pictures, DSS means that the space looks like an inhomogeneous
cone - a cone that was deformed in a periodic manner, see figure
\ref{cone-illus}. In formulae, DSS means that there is a
transformation on space-time $x \to x'=f(x)$ such that the
solution is invariant up to rescaling \bea
 g'_{\mu\nu}(x)=e^{-2\Delta}\, g_{\mu\nu}(x) \non
 \Phi(x')=\Phi(x) + \kappa_\phi ~,\label{DSS} \eea
 where $g'$ is the induced metric $g_{\mu\nu} \to
g'_{\mu\nu}(x)=f^*(g_{\mu\nu})(x')$. The scalar field allows for a
shift constant $\kappa_\phi$ consistent with DSS, but in the
Choptuik critical solution, $\kappa_\phi = 0$ ``for unknown
reasons'' \cite{GundlachRev}.
 In standard coordinates $f$ is conveniently given by $f(\tr,\tlt)=(e^{-\Delta}\,
 \tr,\, e^{-\Delta}\, \tlt)$, or equivalently $\trho \to e^{-\Delta}\, \trho$. The
log-period $\Delta$ was numerically measured in 4d to be  \be
 \Delta(d=4) \simeq 3.45 ~.\ee
 See table \ref{table2} for a collection of these and
other measurements.

Continuous Self Similarity (CSS) would mean for a geometry to have a
transformation satisfying (\ref{DSS}) for all $\Delta$, and that
$\kappa_\phi = const~ \Delta$.

Universality is a consequence of $Z_*$ being an attractor on the
co-dimension 1 surface in phase space. Actually, for DSS the
attractor is a ``limit-cycle''.

Discovering ``echoing'' required special determination and
high-quality numerics, since it was necessary to follow the
solution for several periods, each period lasting a factor of
$e^\Delta \simeq 30$.

\presub \emph{Analogy with second order phase transitions}. The
two phenomena of critical exponents and scale invariance are the
hallmark properties of second order phase transitions in field
theory (for instance, the Curie transition in ferromagnetic
materials, or the liquid-gas critical point). There the critical
point is described by a Conformal Field Theory that looks the same
on all scales. If one deviates from the conformal point the power
law behavior of correlation functions is replaced by a finite
correlation length. Critical exponents appear which are related to
the dimensions of certain operators in the conformal point.
Moreover, reaching the CFT requires tuning some control
parameters, much as the Choptuik solution is gotten after
fine-tuning the initial conditions. Despite this strong analogy
with CFT, a CFT is not known to appear in Choptuik scaling. Given
the modern perspective of Holography via the AdS/CFT
correspondence and the duality of 2d gravity and matrix models, it
is natural to predict that the critical solution is dual to a
large $N$ conformal matrix model.

\presub {\bf Other results}. To date there has been little
analytic understanding of $\gamma,\, \Delta$. The main information
added is the relation \be
 \gamma=1/\lambda_0 \label{gamma-lambda} \ee
 where $\lambda_0$ is the unique negative eigenvalue of the
critical solution. The reason that there is such a unique
eigenvalue is that $Z_*$ being an attractor throughout the co-dim
1 black-hole-threshold ``wall'' is stable to all perturbations
within the wall (stability means positive eigenvalues), but is
unstable against a deviation outside of it.

The point $O$ is a naked singularity. $O$ is singular as an
immediate consequence of the scaling symmetry - as we get closer
to $O$ the curvature scales up and is unbounded in its
neighborhood. Moreover, there is no horizon yet as we are at the
threshold of black hole formation\footnote{However, it is not
implied that the Cauchy horizon is met by a static asymptotic
observer at finite time.}. This teaches us that naked
singularities are generic at co-dim 1 in phase space, and the
celebrated Cosmic Censorship conjecture must be amended to read
``no naked singularities will form for ``reasonable'' \emph{and
generic} initial conditions''.

The behavior of $M_{BH}$ around $p_*$ gets a sub-leading periodic
correction in DSS \be
 \log(M_{BH}) = \gamma\, \log(p-p_*) + c + f_{GHP}\( \gamma\, \log(p-p_*)+c \) ~,\ee
where $f_{GHP}$ is a universal function (our notation, GHP, stands
for Gundlach-Hod-Piran \cite{Gundlach96,HodPiran}) with period
$\Delta$, while $c$ depends on initial conditions.

Another peculiar phenomenon is that when one inspects the metric
alone, without the scalar field, one finds that the log-frequency
doubles, namely the log-period for the metric is $\Delta/2$. For
the scalar field, on the other hand, only odd frequencies are
present.

\section{The correspondence}
\label{correspondence}

In this section the central claim is stated: \\
\underline{\bf Claim}: The critical merger solution in $D$
dimensions corresponds after a double analytic continuation to a
variant of the critical Choptuik solution in $d=D-1$ dimensions but
with different b.c.: time reversal symmetry replaces axis
regularity.

In order to prove this claim we first demonstrate that the actions
are the same up to a double analytic continuation, and therefore
the equations of motions are identical. Then we analyze the
boundary conditions to show that the solutions are identical as
well.

\subsection{The action}

We first verify that both actions are defined in the same
dimension and with the same matter content, and then we proceed to
consider their form. Both actions are essentially 2d once symmetry
is accounted for: in spherical collapse the two essential
coordinates are the Lorentzian $(\tr,\tlt)$, while in the merger
they are the Euclidean $(r,z)$. The matter content in $d$
dimensional spherical collapse is $d$ dimensional metric plus a
scalar field $\Phi$, which is exactly the matter content of
$D=d+1$ gravity in the merger system once a dimensional reduction
over the time coordinate is performed (the Kaluza-Klein vector
field vanishes due to time reversal symmetry).

The precise form of the action is identical as well, since the
scalar obtained from dimensional reduction is minimally coupled,
exactly as in (\ref{choptuik-action}). In order to exhibit the
precise relation between the fields we proceed to perform this
dimensional reduction in the standard way \be ds^2_{D}=e^{2\, A}\,
dt^2 + \widetilde{ds}^2_{d} \label{dim-reduct-ansatz} \ee the
action is $S=1/(16 \pi G_d)\, \int \sqrt{g_d} d^dx\; e^A\;
\widetilde{R}_d$
 where $G_d=G_D/L$
 and after Weyl rescaling \be
 ds^2_d=e^{2\,A/(d-2)}\, \widetilde{ds}^2_{d} \ee one obtains \be
 S =1/(16 \pi G_d)\, \int \sqrt{g_d} d^dx\, \( R_d -{d-1 \over d-2}\, (\del A)^2
 \) \label{Weyl-rescaling-ansatz} \ee
 finally one may rescale $A$ to obtain a canonically normalized
$\Phi$, \be \Phi=\sqrt{{2(d-1) \over (d-2)}} ~ A~,
\label{Phi-normalization} \ee
 yielding the action for spherical collapse
(\ref{choptuik-action}), up to the different signatures (and a
signature related sign). Alternatively, in the spherical collapse
we may consider $\Phi$, the dilaton, to arise from a dimensional
reduction over an additional dimension, which we parametrize by
$\tz$ (and is analogous with $t$ in (\ref{dim-reduct-ansatz}) ).

Moreover, the isometries of spherical collapse and of the merger
are identical: \\ $SO(D-3)_\Omega \times U(1)_t \equiv
SO(d-2)_\Omega \times U(1)_{\tz}$.

More explicitly, it is standard to give the ansatz for spherical
collapse as \bea ds^2_d &=& -\alpha(\tr,\tlt)^2\, d\tlt^2+
a(\tr,\tlt)^2\, d\tr^2 + \tr^2\, d\Omega^{~2}_{d-2} \non
 \Phi &=& \Phi(\tr,\tlt) \label{Choptuik-ansatz} ~.\eea
 When compared with (\ref{general-ansatz},\ref{dim-reduct-ansatz})
we see that in the standard ansatz the gauge freedom is used to set
\bea
 e^{C+{1 \over d-2}\, A} &\to& \tr \non
 ds^2_{\tlt,\tr} &\to& -\alpha(\tr,\tlt)^2\, d\tlt^2+ a(\tr,\tlt)^2\,
 d\tr^2~. \eea

\subsection{Boundary conditions}
\label{BC-subsection}

In order to fully define the solutions we must supply boundary
conditions. For the Choptuik solution the b.c. are \bi
 \item In the scaling direction ($\trho$ - see figure \ref{Choptuik-coord}
the evolution leads to the critical solution, as one evolves
towards the singularity due to the solution's attractor nature.
 It was shown \cite{Gundlach95,Gundlach96} that the attractor mechanism could be
 replaced by periodic b.c.
\item In the ``tangential'' direction (such as $\tr$ for Choptuik)
the b.c. are regularity on the $r=0$ axis and analyticity on the
(past) horizon.
 \ei

When the global properties of the Choptuik solutions were analyzed
\cite{Gundlach96,GarciaGundlachGlobal} it was found that the
standard solution in the ``past'' patch  could be smoothly
continued into the  ``outer'' patch (see figure
\ref{Choptuik-coord}) for the definition of these patches)
delineated by the past and future horizons. However, the
continuation of the standard Choptuik solution is not analytic on
the future horizon. This raises the possibility to define a
different ``tangential'' b.c.: analyticity on both future and past
horizons in the outer patch, or alternatively time reversal
symmetry and horizon regularity. We term the solution obtained
with these b.c. ``time-symmetric Choptuik'' or ``TS-Choptuik'' for
short.

For the critical merger solution the b.c. are very similar \bi
 \item In the scaling direction ($\rho$) we expect an attractor at
criticality, or equivalently self-similarity and periodicity.
 \item In the ``tangential'' direction, $\chi$ or $z$, the boundary
 conditions are reflection symmetry $z \to -z$ or alternatively
 $\chi \to \pi - \chi$, together with regularity on the horizon,
 namely that as $\chi \to 0$ there is no conical deficit angle in
 the Euclidean geometry.  \ei

We see that the Choptuik and merger critical solutions have the
same b.c. in the scaling direction, but different ones in the
tangential direction: Choptuik has axis regularity while the
merger has time reflection symmetry.

Thus, \emph{the merger becomes the TS-Choptuik} after the
following analytic continuation \bea
 z &\leftrightarrow& i\, \tlt \non
 t &\leftrightarrow& i\, \tz \non
 r &\leftrightarrow& \tr ~. \label{time-sym-corres} \eea
While the second analytic continuation is trivial as the fields do
not depend on this coordinate, the first is non-trivial as it
involves an essential coordinate ($z$ or $\tlt$) and it is crucial
that the fields are even in that coordinate in order to retain
reality after analytic continuation. For example for the scalar
field $\Phi(-\tlt)=\Phi(\tlt)$ and hence $\Phi$ is a function of
$\tlt^2$ (namely, there exists some real analytic function
$\hat{\Phi}$ such that $\Phi(\tlt)=\hat{\Phi}(\tlt^2)$) and analytic
continuation sends $\tlt^2 \to -\tlt^2$ keeping the function real
(namely $\hat{\Phi}(\tlt^2) \to \hat{\Phi}(-\tlt^2)$).

\vspace{.5cm} Comments: \begin{enumerate}

 \item In all cases there are two
boundaries in the tangential direction: axis and horizon. So far
we paid most of the attention to the axis, while the boundary
conditions on the horizon were always ``regularity''. Note
however, that ``regularity'' of the horizon has two different
meanings: in the Lorentzian case it means that one can pass
smoothly to Kruskal-like coordinates, while in the Euclidean we
have the ``no deficit angle'' boundary condition for the scalar
field that plays the role that $g_{tt}$ has in ordinary static
geometries.

\item I find it likely, though not self-evident that solutions of
``standard'' spherical collapse respect a $\tr \to -\tr$ symmetry.
It is certainly obeyed by smooth spherically symmetric scalar fields
in a flat background, but the extension to curved space-time is not
obvious to me. If this reflection symmetry does indeed exist then
the Choptuik critical solution can be analytically continued via
$\tr \leftrightarrow i\, r, ~
 \tz \leftrightarrow i\, t, ~
 \tlt \leftrightarrow z$
to a variant of the merger where the $z$-reflection b.c. is replaced
by an axis regularity b.c. at $r=0$.

\end{enumerate}

\subsection{An example}

Here I give an explicit example for the correspondence (double
analytic continuation) between a merger metric\footnote{\cite{AKS}
gives significant evidence that this is actually \emph{the} critical
merger solution, namely the attractor.}
 and the corresponding TS-Choptuik metric.

Consider the metric for a cone over $\IS^2 \times \IS^{D-3}$
(\ref{double-cone-metric}). Identifying the coordinates $\chi,\,
t$ as in \cite{TopChange} and working with a Lorentzian metric we
get the merger-type metric \be
 ds^2= d\rho^2+{\rho^2 \over D-2} \[ d\chi^2-\cos^2(\chi)\, dt^2
 +(D-4)\, d\Omega^2_{\IS^{D-3}} \] ~, \ee
 where $\chi$ was chosen in a slightly non-standard way to belong
to the range $-\pi \le \chi \le \pi$ such that the $\IZ_2$
symmetry reflection symmetry acts simply as $\chi \to -\chi$ and
$t$ was identified such that the metric is independent of $t$ and
moreover $g_{tt}$ vanishes at the boundaries of $\chi$ (the
horizon). Now we perform the double analytic continuation
(\ref{time-sym-corres}) appropriate for the time-symmetric case,
where $\chi$ plays the role of $z$ (the coordinate with the
reflection symmetry) and find \be
 ds^2= d\trho^2+{\trho^2 \over D-2} \[ -d\tlt^2+\cosh^2(\tlt)\, d\tz^2
 +(D-4)\, d\Omega^2_{\IS^{D-3}} \] ~. \ee
Finally, performing a dimensional reduction over $\tz$ according
to (\ref{Weyl-rescaling-ansatz}) with \\ $e^{2A}= \trho^2\,
\cosh^2(\tlt)/(d-1)$, $~d=D-1$, and then normalizing $\Phi$
according to (\ref{Phi-normalization}) we get \bea
 ds^2 &=& \( \trho^2\, \cosh^2(\tlt) \over d-1 \)^{1/(d-2)}\,
 \[ d\trho^2+{\trho^2 \over D-2} \( -d\tlt^2 +(D-4)\, d\Omega^2_{\IS^{D-3}} \)\] \non
 \Phi &=& \sqrt{2(d-1) \over d-2} \,
 \( \log(\trho)+\log(\cosh(\tlt))-\half \log(d-1) \) ~.\eea
 We note that the metric in the $\trho,\tlt$
plane (the outer wedge in figure \ref{Choptuik-coord}) is
conformal to the Rindler metric, namely a wedge in 2d Minkowski
space.

\subsection{Cones and GHP oscillations} \label{cones-GHP}

The action and boundary conditions are (continuously) scale
invariant. Therefore it is natural to start by looking for
 continuously self-similar (CSS) solutions, also known as
cones. The most general CSS ansatz is \bea
 ds^2_d &=& e^{2\, B_\rho(\chi)}\, \( d\rho+\rho\, \hat{A}(\chi)\, d\chi \)^2
 + \rho^2\, e^{2\, B_\chi(\chi)}\,d\chi^2
 +\rho^2\, e^{2\, C(\chi)}\, d\Omega_{d-2}^{~2} \non
 \Phi(\rho,\chi) &=& \kappa_\phi\, \rho + \Phi(\chi) \eea
 where all fields $B_\rho, B_\chi, \hat{A}, C, \Phi$ depend only on
$\chi$. Examples of cones include the double-cone
(\ref{double-cone-metric}) and the Roberts solution \cite{Roberts}
\bea
 ds^2 &=& -du\, dv + \tr^2(u,v)\, d\Omega^2 \non
 \tr^2(u,v) &=& \[ (1-p^2)\, v^2 - 2\, v\, u + u^2 \] \non
 \Phi &=& \half \log{(1-p)\,v -u \over (1+p)\, v - u} \label{roberts} ~,\eea
 where $p=1$ is a critical value.

Assuming that the double cone is indeed the critical merger
solution \cite{AKS}, the exponents $s$
(\ref{linearized-exponents}) which appear at the linearized level
can be interpreted as follows. Take (\ref{linearized-exponents})
and perform two substitutions. First substitute $\eps \to
 \delta p:=p-p^*$ for the deviations
from the double cone. Second, replace $\rho \to \rho/\rho_0$ in
order for the expression to be dimensionally correct, and $\rho_0$
will be interpreted as a length scale characteristic of the smooth
cone, for example, $\rho_0^{\,-2}$ could be a measure of its maximal
curvature. The result is \be
 \delta p \sim (\rho/\rho_0)^s \sim \rho_0^{-s} \label{srho0}~.\ee
Therefore \be
 \rho_0 \sim \delta p^{-1/s} .\ee
In the theory of critical collapse an analogous relation defines
the critical exponents $\gamma,\Delta$ \be
 \rho_0 \sim \delta p^{\gamma(1 \pm i\, 2 \pi/\Delta)} \label{GHP} ~,\ee
where $\rho_0^{\,-2}$ is a measure of the maximal curvature above or
below criticality, and $\Delta$ is the log-period of the GHP
oscillations. Normally one writes only the real part of the exponent
$\rho_0 \sim \delta p^{\gamma}$ and (\ref{GHP}) is a compact form
which includes also the GHP oscillations. Moreover for critical
collapse $\Delta$ measured from GHP oscillations is the same as the
log-period of the critical (DSS) solution, $Z_*$.

Comparing (\ref{srho0},\ref{GHP}) we find that $s$ is \emph{a
complex quantity which naturally combines the two scaling
constants} $\gamma,\, \Delta$ through \be
 -{1 \over s}=\gamma\, \( 1 \pm i\, {2\,\pi\, \over \Delta} \)
\label{complexify}~. \ee Therefore \bea
 \gamma &=& -\Re\({1 \over s}\)  \label{gamma-pred1}
 \\
 {2 \pi \over \Delta} &=& {\Im(1/s) \over \Re(1/s)} ~. \label{Delta-pred1}
 \eea

Combining (\ref{gamma-pred1},\ref{Delta-pred1}) with the explicit
expressions for $s$ (\ref{linearized-exponents}) we may predict
off-critical oscillations for the merger at $D<10$, with the
following critical exponents \bea
 \gamma &=& {1 \over 4} \label{gamma-pred2}
 \\
 {2 \pi \over \Delta} &=&  \sqrt{{10-D \over D-2}}~,\label{Delta-pred2}
 \eea
while for $D \ge 10$ there are no oscillations and the critical
exponent $\gamma$ becomes \be
 \gamma = -{1 \over s_+}={1 \over 4}\, \(1 + \sqrt{{D-10 \over D-2}} \) ~,\label{gamma-pred3} \ee
where $s_+$ was substituted in (\ref{gamma-pred1}) since it is
leading for large $\rho$.

Equation (\ref{gamma-pred1}) may be compared with
(\ref{gamma-lambda}), the well-known connection in the theory of
critical collapse.

\section{Consequences and indications}
\label{consequences}

The correspondence outlined in the previous section suggests certain
new directions for numerical work. \bi

%%%%%%%%%%%%%%%%%%%%%%%%%%%%%%%%%%%%%%%%%%%%%%%%%%%%
\item {\bf Predictions for off-critical oscillations and for the
scaling constants of the merger for $D<10$}.

Off-critical oscillations in the merger system are predicted to be
analogous with GHP oscillations \cite{Gundlach96,HodPiran} in
critical collapse. The predicted values of the two scaling constants
are given in (\ref{gamma-pred2},\ref{Delta-pred2}).

\noindent \emph{Suggested numerical experiment}. Measure the scaling
constants $\gamma, \Delta$ from off-critical merger solutions for
various dimensions $D<10$, improving on the pioneering work of
\cite{Wiseman1,KolWiseman}. According to (\ref{GHP}) $\gamma$ may be
defined through the maximal curvature as one goes off criticality,
exactly as in Choptuik scaling, and $\Delta$ is defined such that
the period in $\log\,\delta p$ is $\Delta/\gamma$.

%%%%%%%%%%%%%%%%%%%%%%%%%%%%%%%%%%%%%%%%%%%%%%%%%%%%
\item {\bf Critical dimension $D^*=10$ for merger, and a
prediction for $\gamma$}.

For $D \ge D^*=10$ GHP oscillations are predicted to cease to exist
as a consequence of the property described around
(\ref{linearized-exponents}). The prediction for $\gamma$ becomes
(\ref{gamma-pred3}).

\noindent \emph{Suggested numerical experiment}. Seek the
off-critical behavior for the merger in dimensions $D \ge D^*=10$.
%%%%%%%%%%%%%%%%%%%%%%%%%%%%%%%%%%%%%%%%%%%%%%%%%%%
\item {\bf Possible similarity between Choptuik scaling and the
merger}.

The merger and Choptuik systems were shown to have the same action
after analytic continuation, but different boundary conditions. At
the time when this paper was conceived, the data regarding the
Choptuik scaling constants in various dimensions was scarce, and
seemed to compare surprisingly well with the prediction for the
merger (\ref{Delta-pred2}):
the predicted $\Delta$ for merger is $\Delta|_{D=5}=\Delta|_{D=7} =4
\pi /\sqrt{15} \simeq 3.24$ while the available $\Delta$'s for
critical collapse were $\Delta(4d) \simeq 3.45$, $\Delta(6d)=3.03$
\cite{6dChoptuik} and $d=D-1$ (the available data at the time is
summarized in table \ref{table2}).
%, as we describe below.
That led to the suggestion that perhaps the change in boundary
conditions would affect the scaling constants only weakly, making
the Choptuik constants always close to
(\ref{gamma-pred1},\ref{Delta-pred2},\ref{gamma-pred3}). Moreover it
suggested the possibility that critical dimensions which are known
for the merger system \cite{TopChange,SorkinD*} will appear in
Choptuik scaling as well.

\begin{table}[t]
\centering
\begin{tabular}{l||c|c}
 & {\bf $\Delta$} & {\bf $\gamma$} \\
 \hline \hline
\room {\bf 4d} & 3.45 & 0.374 \\
\hline
\room {\bf 6d} & 3.03 &  0.424 \\
\end{tabular}
\caption[]{Scaling constants for the Choptuik critical collapse in
4d \cite{Choptuik} and 6d \cite{6dChoptuik} which were available at
the time this paper was conceived. $\Delta$ is the log-period, and
$\gamma$ is the scaling exponent. By now a strikingly precise
determination of $\Delta(d=4)$ is available: $\Delta(d=4) \simeq
3.445452402(3)$ \cite{GarciaGundlachGlobal}. In higher dimensions
$\gamma$ is defined such that $M \simeq (p-p_*)^{\gamma\,(d-3)}$,
namely, $(p-p_*)^\gamma$ has length dimension 1. For newer data in
various dimensions see \cite{SorkinOren}.}  \label{table2}
\end{table}

Inspired by these ideas and motivated by the apparent success of the
estimates in 4d, 6d Sorkin  and Oren set out to measure the scaling
constants $\gamma,\, \Delta$ for critical collapse in $d \le 11$
\cite{SorkinOren} (see \cite{KimMoonYee-dimensional,Birukou:2002kk}
for previous attempts). They succeeded and their interesting results
indicate that (\ref{Delta-pred2}) is \emph{not} a good estimator for
the Choptuik $\Delta$ and the good agreement in certain dimensions
which was described above should be considered a coincidence. The
results for $\gamma$ are not very close either. This is not
surprising
 in view of the differences between the systems both in b.c. and in
the DSS vs. CSS nature of the critical solution.

Regarding a critical dimension their results are less conclusive.
No phase transition to CSS was observed up to $11d$, but there are
some indications for extrema of the scaling constants (as a
function of dimension) shortly above 11d. Note that the large $d$
simulations are impeded by a growingly singular behavior at the
$\tr=0$ axis.

\ei

\vspace{0.5cm}
 \noindent {\bf Acknowledgements}

The author is indebted to Nissan Itzhaki for important discussions
and participation in the early part of this work. It is a pleasure
to thank Evgeny Sorkin for discussions, and him and Yonatan Oren
for sharing the results of \cite{SorkinOren} prior to publication.
I also wish to thank members of the theoretical physics group at
the Hebrew University for various discussions.

It is a great pleasure to thank my brother, Boaz Kol, for a
discussion on self-organized criticality, a central topic of his
PhD thesis.

BK is supported in part by The Israel Science Foundation (grant no
228/02) and by the Binational Science Foundation BSF-2002160.


\newpage

\begin{thebibliography}{99}

\bibitem{Choptuik}
M.~W.~Choptuik, ``Universality and scaling in gravitational
collapse of a massless scalar field,'' Phys.\ Rev.\ Lett.\  {\bf
70}, 9 (1993).
%%CITATION = PRLTA,70,9;%%

\bibitem{GundlachRev}
C.~Gundlach, ``Critical phenomena in gravitational collapse,''
Phys.\ Rept.\  {\bf 376}, 339 (2003) [arXiv:gr-qc/0210101].
%%CITATION = GR-QC 0210101;%%

\bibitem{TopChange}
  B.~Kol,
  ``Topology change in general relativity and the black-hole black-string
  transition,''
  JHEP {\bf 0510}, 049 (2005)
  [arXiv:hep-th/0206220].
  %%CITATION = HEP-TH 0206220;%%


\bibitem{review}
  B.~Kol,
  ``The phase transition between caged black holes and black strings: A
  review,''
  Phys.\ Rept.\  {\bf 422}, 119 (2006)
  [arXiv:hep-th/0411240].
  %%CITATION = HEP-TH 0411240;%%

\bibitem{KudohWiseman2}
  H.~Kudoh and T.~Wiseman,
  ``Connecting black holes and black strings,''
  Phys.\ Rev.\ Lett.\  {\bf 94}, 161102 (2005)
  [arXiv:hep-th/0409111].
  %%CITATION = HEP-TH 0409111;%%


\bibitem{GIR}
D.~Gaiotto, N.~Itzhaki and L.~Rastelli, ``Closed strings as
imaginary D-branes,'' Nucl.\ Phys.\ B {\bf 688}, 70 (2004)
[arXiv:hep-th/0304192].
%%CITATION = HEP-TH 0304192;%%

\bibitem{MSY}
A.~Maloney, A.~Strominger and X.~Yin, ``S-brane thermodynamics,''
JHEP {\bf 0310}, 048 (2003) [arXiv:hep-th/0302146].
%%CITATION = HEP-TH 0302146;%%

\bibitem{JMS}
G.~Jones, A.~Maloney and A.~Strominger, ``Non-singular solutions
for S-branes,'' Phys.\ Rev.\ D {\bf 69}, 126008 (2004)
[arXiv:hep-th/0403050].
%%CITATION = HEP-TH 0403050;%%

\bibitem{Frolov4}
A.~V.~Frolov, ``Continuous self-similarity breaking in critical
collapse,'' Phys.\ Rev.\ D {\bf 61}, 084006 (2000)
[arXiv:gr-qc/9908046].
%%CITATION = GR-QC 9908046;%%

\bibitem{Frolov2}
A.~V.~Frolov, ``Self-similar collapse of scalar field in higher
dimensions,'' Class.\ Quant.\ Grav.\  {\bf 16}, 407 (1999)
[arXiv:gr-qc/9806112].
%%CITATION = GR-QC 9806112;%%

\bibitem{Roberts}
M.~D.~Roberts, ``Scalar field counterexamples to the cosmic
censorship hypothesis,'' Gen.\ Rel.\ Grav.\  {\bf 21}, 907 (1989).
%%CITATION = GRGVA,21,907;%%

\bibitem{CLOPPV}
M.~W.~Choptuik, L.~Lehner, I.~Olabarrieta, R.~Petryk, F.~Pretorius
and H.~Villegas, ``Towards the final fate of an unstable black
string,'' Phys.\ Rev.\ D {\bf 68}, 044001 (2003)
[arXiv:gr-qc/0304085].
%%CITATION = GR-QC 0304085;%%


\bibitem{Gundlach96}
C.~Gundlach, ``Understanding critical collapse of a scalar
field,'' Phys.\ Rev.\ D {\bf 55}, 695 (1997)
[arXiv:gr-qc/9604019].
%%CITATION = GR-QC 9604019;%%

\bibitem{GarciaGundlachGlobal}
J.~M.~Martin-Garcia and C.~Gundlach, ``Global structure of
Choptuik's critical solution in scalar field collapse,'' Phys.\
Rev.\ D {\bf 68}, 024011 (2003) [arXiv:gr-qc/0304070].
%%CITATION = GR-QC 0304070;%%

\bibitem{HodPiran}
S.~Hod and T.~Piran, ``Fine-structure of Choptuik's mass-scaling
relation,'' Phys.\ Rev.\ D {\bf 55}, 440 (1997)
[arXiv:gr-qc/9606087].
%%CITATION = GR-QC 9606087;%%

\bibitem{Gundlach95}
C.~Gundlach, ``The Choptuik space-time as an eigenvalue problem,''
Phys.\ Rev.\ Lett.\  {\bf 75}, 3214 (1995) [arXiv:gr-qc/9507054].
%%CITATION = GR-QC 9507054;%%

\bibitem{Wiseman1}
T.~Wiseman, ``Static axisymmetric vacuum solutions and non-uniform
black strings,'' Class.\ Quant.\ Grav.\  {\bf 20}, 1137 (2003)
[arXiv:hep-th/0209051].
%%CITATION = HEP-TH 0209051;%%

\bibitem{KolWiseman}
B.~Kol and T.~Wiseman, ``Evidence that highly non-uniform black
strings have a conical waist,'' Class.\ Quant.\ Grav.\  {\bf 20},
3493 (2003) [arXiv:hep-th/0304070].
%%CITATION = HEP-TH 0304070;%%

\bibitem{6dChoptuik}
D.~Garfinkle, C.~Cutler and G.~C.~Duncan, ``Choptuik scaling in
six dimensions,'' Phys.\ Rev.\ D {\bf 60}, 104007 (1999)
[arXiv:gr-qc/9908044].
%%CITATION = GR-QC 9908044;%%

\bibitem{SorkinD*}
E.~Sorkin, ``A critical dimension in the black-string phase
transition,'' Phys.\ Rev.\ Lett.\  {\bf 93}, 031601 (2004)
[arXiv:hep-th/0402216].
%%CITATION = HEP-TH 0402216;%%

\bibitem{SorkinOren}
  E.~Sorkin and Y.~Oren,
  ``On Choptuik's scaling in higher dimensions,''
  Phys.\ Rev.\ D {\bf 71}, 124005 (2005)
  [arXiv:hep-th/0502034].
  %%CITATION = HEP-TH 0502034;%%

\bibitem{ItzhakiPrivat}
 N.~Itzhaki and B.~Kol, private discussion.

\bibitem{AKS}
  V.~Asnin, B.~Kol and M.~Smolkin,
   ``Analytic Evidence for Continuous Self Similarity of the Critical Merger
  Solution,''
  arXiv:hep-th/0607129.
  %%CITATION = HEP-TH 0607129;%%

\bibitem{KimMoonYee-dimensional}
H.~C.~Kim, S.~H.~Moon and J.~H.~Yee, ``Dimensional dependence of
black hole formation in scalar field  collapse,'' JHEP {\bf 0202},
046 (2002) [arXiv:gr-qc/0108071].
%%CITATION = GR-QC 0108071;%%


\bibitem{Birukou:2002kk}
M.~Birukou, V.~Husain, G.~Kunstatter, E.~Vaz and M.~Olivier,
``Scalar field collapse in any dimension,'' arXiv:gr-qc/0201026.
%%CITATION = GR-QC 0201026;%%

\end{thebibliography}
\end{document}